\documentclass[conference]{IEEEtran}

\usepackage{algorithm,algorithmic}
\usepackage{amsmath,amssymb,mathrsfs}
\usepackage{array}
\usepackage{balance}
\usepackage{cite}
\usepackage{epsfig}
\usepackage{easyReview}
\usepackage{fancyhdr}
\usepackage{float}
\usepackage{glossaries}
\usepackage{graphicx,color}
\usepackage{hyperref}
\usepackage{mathtools}
\usepackage{multirow}
\usepackage{psfrag}
\usepackage{stfloats}
\usepackage{subfigure}
\usepackage{url}

\newcommand{\figref}[1]{Fig. \ref{#1}}

\renewcommand{\eqref}[1]{(\ref{#1})}

\definecolor{sblue}{RGB}{0,51,120}
\definecolor{sred}{RGB}{139,0,139}
\definecolor{sg}{RGB}{46,139,87}

\makeglossaries
\newacronym[description=Additive white Gaussian noise]{awgn}{AWGN}{additive white Gaussian noise}
\newacronym[description=Approximate message passing]{amp}{AMP}{approximate message passing}
\newacronym[description={\em a posteriori} probability]{app}{APP}{{\em a posteriori} probability}
\newacronym[description=Base station]{bs}{BS}{base station}
\newacronym[description=Base station sleeping ]{bss}{BSS}{base station sleeping }
\newacronym[description=Belief propagation]{bp}{BP}{belief propagation}
\newacronym[description=Binary phase shift keying]{bpsk}{BPSK}{binary phase shift keying}
\newacronym[description=Bit error rate]{ber}{BER}{bit-error-rate}
\newacronym[description=Block error rate]{bler}{BLER}{block error rate}
\newacronym[description=Central limit theorem]{clt}{CLT}{central limit theorem}
\newacronym[description=Channel state information ]{csi}{CSI}{channel state information }
\newacronym[description=Closest vector problem]{cvp}{CVP}{closest vector problem}
\newacronym[description=Code division multiple access]{cdma}{CDMA}{code division multiple access}
\newacronym[description=Distributed linear data fusion]{dldf}{DLDF}{distributed linear data fusion}
\newacronym[description=European Cooperation in Science and Technology]{cost}{COST}{Cooperation in Science and Technology}
\newacronym[description=Coordinated multi-point ]{CoMP}{CoMP}{coordinated multi-point }
\newacronym[description=Correlation-based stochastic model]{cbsm}{CBSM}{correlation-based stochastic model}
\newacronym[description=Cumulative distribution function]{cdf}{CDF}{cumulative distribution function}
\newacronym[description=Degrees of freedom]{dof}{DoF}{degrees of freedom}
\newacronym[description=Element-based lattice reduction]{elr}{ELR}{element-based lattice reduction}
\newacronym[description=Extremely-large aperture array]{elaa}{ELAA}{extremely-large aperture array}
\newacronym[description=Fifth-generation]{5g}{5G}{fifth-generation}
\newacronym[description=Fixed-complexity sphere decoder]{fcsd}{FCSD}{fixed-complexity sphere decoder}
\newacronym[description=Forward error corrrection]{fec}{FEC}{forward error correction}
\newacronym[description=Free space path loss]{fspl}{FSPL}{free space path loss}
\newacronym[description=Global system for mobile communication]{gsm}{GSM}{global system for mobile communication}
\newacronym[description=Geometry-based stochastic model]{gbsm}{GBSM}{geometry-based stochastic model}
\newacronym[description=Hermite-Korkin-Zolotarev]{hkz}{HKZ}{Hermite-Korkin-Zolotarev}
\newacronym[description=Independent and identically distributed]{iid}{i.i.d.}{independent and identically distributed}
\newacronym[description=Integer least-squares]{ils}{ILS}{integer least-squares}
\newacronym[description=International Telecommunication Union Radiocommunication Sector ]{itu-r}{ITU-R}{International Telecommunication Union Radiocommunication Sector}
\newacronym[description=Large system behaviour]{lsb}{LSB}{large system behaviour}
\newacronym[description=Lattice reduction]{lr}{LR}{lattice reduction}
\newacronym[description=Lenstra-Lenstra-Lov\'{a}sz]{lll}{LLL}{Lenstra-Lenstra-Lov\'{a}sz}
\newacronym[description=Likelihood ascent search]{las}{LAS}{likelihood ascent search}
\newacronym[description=Line-of-slight]{los}{LoS}{line-of-slight}
\newacronym[description=List sphere decoder]{lsd}{LSD}{list sphere decoder}
\newacronym[description=Linear minimum mean square error]{lmmse}{LMMSE}{linear minimum mean square error}
\newacronym[description=Log-likelihood ratio]{llr}{LLR}{log-likelihood ratio}
\newacronym[description=Long-term evolution ]{lte}{LTE}{long-term evolution}
\newacronym[description=Low density parity check]{ldpc}{LDPC}{low density parity check}
\newacronym[description=Massive machine type communications]{mmtc}{mMTC}{massive machine type communications}
\newacronym[description=Maximum {\em a posteriori}]{map}{MAP}{maximum {\em a posteriori}}
\newacronym[description=Maximum likelihood sequence detection]{mlsd}{MLSD}{maximum likelihood sequence detection}
\newacronym[description=Multiple-input multiple-output]{mimo}{MIMO}{multiple-input multiple-output}
\newacronym[description=massive multiple-input multiple-output]{mmimo}{mMIMO}{massive multiple-input multiple-output}
\newacronym[description=Matched filter]{mf}{MF}{matched filter}
\newacronym[description=Mean square error]{mse}{MSE}{mean square error}
\newacronym[description=Minimum mean square error]{mmse}{MMSE}{minimum mean square error}
\newacronym[description=Mobile and wireless communications Enablers for the Twenty-twenty Information ]{metis}{METIS}{Mobile and wireless communications Enablers for the Twenty-twenty Information}
\newacronym[description=Non-line-of-sight]{nlos}{NLoS}{non-LoS}
\newacronym[description=One dimensional]{1d}{1-D}{one dimensional}
\newacronym[description=Orthogonality defect]{od}{OD}{orthogonality defect}
\newacronym[description=Pairwise error probability]{pep}{PEP}{pairwise error probability}
\newacronym[description=Parallel interference cancellation]{pic}{PIC}{parallel interference cancellation}
\newacronym[description=Probabilistic data association]{pda}{PDA}{probabilistic data association}
\newacronym[description=Probability distribution function]{pdf}{PDF}{probability density function}
\newacronym[description=Probability mass function]{pmf}{PMF}{probability mass function}
\newacronym[description=Quadrature amplitude modulation]{qam}{QAM}{quadrature amplitude modulation}
\newacronym[description=Quadrature phase shift keying]{qpsk}{QPSK}{quadrature phase shift keying}
\newacronym[description=Receiver-side channel state information]{rcsi}{R-CSI}{receiver-side channel state information}
\newacronym[description=Received signal strength]{rss}{RSS}{received signal strength}
\newacronym[description=Semidefinite relaxation]{sdr}{SDR}{semidefinite relaxation}
\newacronym[description=Seysen's algorithm]{sa}{SA}{Seysen's algorithm}
\newacronym[description=Signal-to-interference-plus-noise ratio]{sinr}{SINR}{signal-to-interference-plus-noise ratio}
\newacronym[description=Signal to interference ratio]{sir}{SIR}{signal to interference ratio }
\newacronym[description=Signal-to-noise ratio]{snr}{SNR}{signal-to-noise ratio}
\newacronym[description=Single antenna interference cancellation]{saic}{SAIC}{single antenna interference cancellation}
\newacronym[description=Single input single output]{siso}{SISO}{single input single output}
\newacronym[description=Singular value decomposition ]{svd}{SVD}{singular value decomposition }
\newacronym[description=Sixth-generation mobile networks]{6g}{6G}{sixth-generation}
\newacronym[description=Sphere decoder]{sd}{SD}{sphere decoder}
\newacronym[description=Space-time codes]{stc}{STC}{space-time codes}
\newacronym[description=State-of-the-art]{sota}{SoTA}{state-of-the-art}
\newacronym[description=Successive interference cancellation]{sic}{SIC}{succesive interference cancellation}
\newacronym[description=Symbol error rate]{ser}{SER}{symbol-error-rate}
\newacronym[description=Tabu search]{ts}{TS}{tabu search}
\newacronym[description=Three-dimensional]{3d}{3-D}{three-dimensional}
\newacronym[description=The 3rd Generation Partnership Project]{3gpp}{3GPP}{the 3rd Generation Partnership Project}
\newacronym[description=Two-dimensional]{2d}{2-D}{two-dimensional}
\newacronym[description=Uniform linear array]{ula}{ULA}{uniform linear array}
\newacronym[description=Urban micro]{umi}{UMi}{urban micro}
\newacronym[description=User equipment]{ue}{UE}{user equipment}
\newacronym[description=Vector error rate]{ver}{VER}{vector error rate}
\newacronym[description=Vertical Bell Labs layered space-time]{vblast}{V-BLAST}{vertical Bell Labs layered space-time}
\newacronym[description=Visibility region]{vr}{VR}{visibility region}
\newacronym[description=Widely linear]{wl}{WL}{widely linear}
\newacronym[description=Widely linear zero forcing]{wlzf}{WLZF}{widely linear zero forcing}
\newacronym[description=Wide-sense stationary uncorrelated scattering]{wssus}{WSSUS}{wide-sense stationary uncorrelated scattering}
\newacronym[description=Wireless World Initiative New Ratio]{winner}{WINNER}{Wireless World Initiative New Ratio}
\newacronym[description=Zero forcing]{zf}{ZF}{zero forcing}
\newacronym[description=Zero mean complex circularly symmetric]{zmccs}{ZMCCS}{zero mean complex circularly symmetric}
\newacronym[description=Access Point]{ap}{AP}{access point}

\ifCLASSINFOpdf
\else
\fi

\begin{document}
\title{\huge Importance-Aware Source-Channel Coding for Multi-Modal Task-Oriented Semantic Communication}
\author{ Yi Ma, Chunmei Xu, Zhenyu Liu, Siqi Zhang,  and Rahim Tafazolli\\
	{6GIC, Institute for Communication Systems, University of Surrey, Guildford, UK, GU2 7XH}\\
	{ Emails: (y.ma, chunmei.xu, zhenyu.liu, s.zhang, r.tafazolli)@surrey.ac.uk }
}

\markboth{}%
{Shell \MakeLowercase{\textit{et al.}}: Bare Demo of IEEEtran.cls for IEEE Journals}
\maketitle

\begin{abstract}
This paper explores the concept of information importance in multi-modal task-oriented semantic communication systems, emphasizing the need for high accuracy and efficiency to fulfill task-specific objectives. 
At the transmitter, generative AI (GenAI) is employed to partition visual data objects into semantic segments, each representing distinct, task-relevant information. These segments are subsequently encoded into tokens, enabling precise and adaptive transmission control.
Building on this framework, we present importance-aware source and channel coding strategies that dynamically adjust to varying levels of significance at the segment, token, and bit levels. 
The proposed strategies prioritize high fidelity for essential information while permitting controlled distortion for less critical elements, optimizing overall resource utilization. 
Furthermore, we address the source-channel coding challenge in semantic multiuser systems, particularly in multicast scenarios, where segment importance varies among receivers. To tackle these challenges, we propose solutions such as rate-splitting coded progressive transmission, ensuring flexibility and robustness in task-specific semantic communication.
\end{abstract}

\section{Introduction}
Generative Artificial Intelligence (GenAI) is revolutionizing the fundamental principles of communication systems, including the representation of information (source coding), its protection against errors (channel coding), and its efficient transmission across communication channels (resource allocation). Traditional communication paradigms prioritize reliable reconstruction of the original information source at the destination. 
However, with GenAI, the focus shifts towards goal-oriented and task-specific applications, where the end objective drives the design and optimization of the communication process.

In source coding, traditional methods aim to compress information without loss (e.g., entropy coding) or with acceptable distortion (e.g., lossy compression for audio or video). 
These approaches adhere to Shannon’s coding theorems, where reconstruction fidelity remains central. 
In contrast, GenAI-based source coding does not necessarily aim for exact recovery of the original content but instead generates task-relevant information. For example, instead of reconstructing an image pixel by pixel, GenAI can generate a lower-dimensional sketch, semantic attributes, or task-specific latent representations that are sufficient for downstream goals such as object recognition or classification. 
This efficiency reduces the communication overhead while retaining the actionable value of the data.

For channel coding, traditional systems employ deterministic error-correction codes (e.g., Polar code) that are optimized to minimize bit error rates or codeword error rates. 
These techniques, while robust, rely on predefined codebooks and operate independently of the content being transmitted. 
GenAI introduces adaptive coding mechanisms that leverage learned, content-aware redundancy. 
By exploiting the structure or semantics of the transmitted data, GenAI-based coding can adapt to channel conditions in real time. 
For instance, critical semantic components of a signal might receive higher protection, ensuring that errors affect task performance minimally, even if exact reconstruction is compromised.

In terms of resource allocation, traditional communication systems optimize resources like bandwidth, power, and spectrum to achieve uniform performance metrics such as data rate, latency, and reliability. 
However, these strategies often treat all transmitted information equally, regardless of its importance. 
GenAI-based resource allocation, on the other hand, adopts a task-specific approach. 
By incorporating the significance of data components, resources can be dynamically prioritized. 
For example, in an edge-assisted autonomous vehicle network, more resources should be allocated to transmitting critical features like road boundaries or moving objects, while deprioritizing background information.

Together, these advancements in source coding, channel coding, and resource allocation shift the paradigm from data-centric communication systems to goal-oriented, semantic-aware frameworks. 
The integration of GenAI enables radios to 'understand' and adapt to both the content and the purpose of the transmitted information, leading to unprecedented efficiency, adaptability, and robustness in emerging communication systems.

In this paper, we explore the concept of information importance in multi-modal task-oriented semantic communication systems, emphasizing the need for accuracy and efficiency in transmitting data to meet task-specific requirements. 
At the transmitter, GenAI is utilized to partition a visual data object into multiple semantic segments, each encapsulating distinct and task-relevant information. These segments are then independently encoded into tokens, providing a flexible and granular approach to transmission control.

Building on this framework, we delve into source and channel coding strategies designed to accommodate varying levels of importance across segment, token, and bit dimensions. We specifically investigate how semantic importance can inform the allocation of coding and transmission resources, ensuring high fidelity for critical information while allowing controlled degradation in less essential components. Furthermore, we address the complexities introduced by semantic multicast systems, where the significance of semantic segments may differ across receivers, necessitating a dynamic, context-aware approach to source-channel coding and resource management.
By incorporating these importance-aware strategies, we demonstrate significant improvements in the efficiency and robustness of semantic communication systems, particularly in task-specific application scenarios.

\section{Semantic Segmentation, Kolmogorov Compression, and Semantic Importance}
We consider a communication system where a transmitter aims to deliver a still visual data object to one or more receivers over wireless  channels. 
Analogous to the approach used in JPEG2000, the transmitter performs source coding by first partitioning the data object into multiple semantic segments and then compressing each segment as part of the encoding process.
However, the system of our interest differs from JPEG2000 in three key aspects:

{\em 1)} Semantic segments differ from the regions of interest (RoIs) used in JPEG2000. 
RoIs are user-defined areas prioritized for processing, compression, or analysis without necessarily interpreting their content \cite{JPEG2000}. 
For example, in JPEG2000, RoIs may encode areas like tumors in medical imaging with higher quality. 
Semantic segments, however, are regions defined based on their meaning or object category (e.g., ``car,'' ``tree," or ``sky"), relying on content recognition and modern foundation models (e.g., \cite{kirillov2023segment}) for tasks like object detection or scene understanding. 
In short, RoIs are not designed and optimized for semantic, task-oriented applications.

{\em 2)} Semantic segments in our system are compressed in a manner distinct from the RoIs used in JPEG2000. 
By leveraging modern foundation models (e.g., \cite{saharia2022,xu2024generative}) and language models (e.g., \cite{deletang2024language}), semantic segments are transformed into compact textual descriptions, guided by the principle of Kolmogorov compression \cite{AK1963,AK1998}. This approach emphasizes semantic meaning over pixel-level accuracy.

To illustrate Kolmogorov compression, consider the binary string:
\begin{equation}\label{eq01}
``\underbrace{101010101010 \cdots 101010101010}_{\text{2 million digits}}"
\end{equation}
This sequence can be succinctly described by a simple computer program:
\begin{displaymath}
\texttt{repeat '10' a million times}.
\end{displaymath}
Here, meaningful units like `repeat', `10', and `times' act as tokens, which serve as the fundamental building blocks of modern AI-driven communication. 
This principle has recently been coined as token communication in~\cite{WenICCC2024}, providing a direct application to the compression of semantic segments.

{\em 3)} Not all semantic segments carry equal importance for the tasks at the destination.
Some segments represent critical information, such as key objects in a scene, and therefore require ultra-high fidelity and reliability during source coding and transmission.
In contrast, less significant segments, such as background details or redundant components, can tolerate greater distortion without substantially affecting the utility of the reconstructed data or the performance of downstream applications. 
This type of importance measure is inherently different from JPEG2000, where RoIs are predefined by the user based on static visual importance. 
In contrast, semantic segments are dynamically assessed based on content recognition and task relevance, providing a more adaptive approach to data compression that aligns with the needs of task-oriented communications.

In the context of multiuser semantic communications, such as semantic multicast \cite{liu2024semcom}, the importance of a semantic segment depends on the specific needs of the receivers. Semantic segments that are of higher relevance to the receivers should be prioritized during data transmission.
For example, the semantic map, which is essential for synthesizing semantic segments, is critical for all receivers. In tasks like visual data reconstruction, semantic segments become significantly less useful without the corresponding semantic map. Therefore, prioritizing the semantic map in resource allocation and transmission is crucial.

Beyond the segment-level importance, token communication also introduces token-level importance. 
Generally speaking, some tokens hold greater importance when \cite{vaswani2017attention}
\begin{itemize}
\item They are shorter tokens, as even a single error can have a disproportionately large impact on their interpretation and the overall message.
\item They function as independent tokens, lacking the contextual cues or structural dependencies that could help infer or correct errors through syntactic or semantic rules.
\item They have less internal redundancy, meaning there is little or no inherent structure to help detect or mitigate errors.
\item Their role is critical to decoding or reconstructing other information, acting as key building blocks or references for understanding subsequent data.
\end{itemize}
To illustrate, we revisit the binary sequence in \eqref{eq01} and its associated simple computer program. 
The token {\tt '10'}  is a short, independent, and minimally redundant unit. 
A error in {\tt '10'} is not semantically detectable and would result in the reconstruction of an entirely incorrect sequence. 
This critical role in accurately reproducing the binary sequence establishes {\tt '10'} as the most important token in the program.

The token {\tt 'a'} is also a compact unit, even shorter than {\tt '10'}, and exhibits minimal redundancy. 
However, unlike {\tt '10'}, {\tt 'a'} is not independent. Its relationship with other tokens enables the semantic decoder to achieve a certain degree of error detection. 
For example, if {\tt 'a'} were mistakenly altered to {\tt 'b'}, the semantic decoder would fail to execute the program, prompting a retransmission request.

In contrast, some tokens are inherently more robust to errors. 
For instance, a minor typo in {\tt 'times'} could be automatically corrected by language models, eliminating the need for retransmission. 
This highlights the varying levels of resilience among tokens, shaped by their inherent structure and semantic context.

\begin{figure}[t!]
    \centering
    \includegraphics[width=0.45\textwidth]{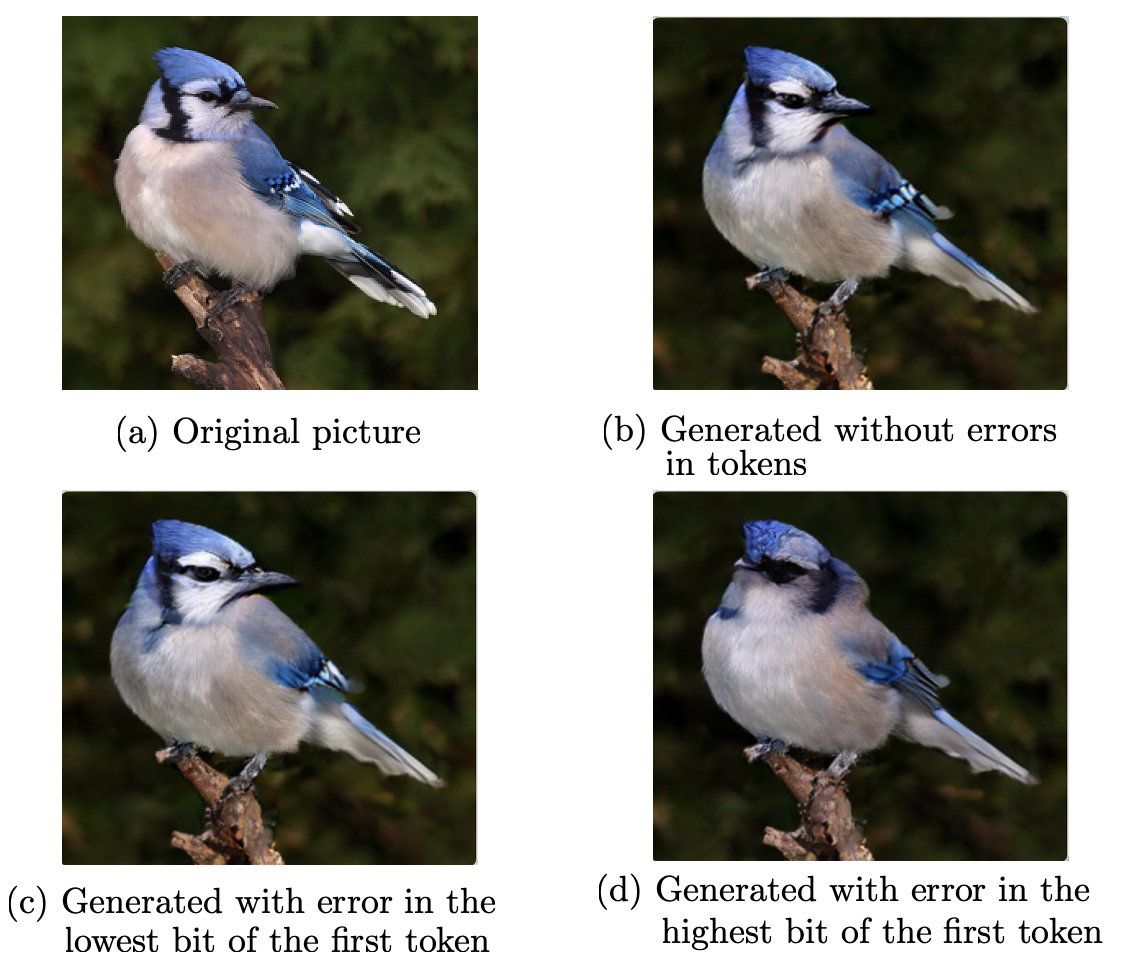} 
    \caption{Kolmogorov compression and reproduction of a picture with or without errors in token communication. MS-SSIM score: (b) $0.598$, (c) $0.541$, (d) $0.539$.}
    \label{fig:bird}
\end{figure}
Kolmogorov compression is not limited to textual transformations; it extends to any form of symbolic representation, including numerical sequences, binary encodings, or even abstract mathematical structures. For example, as demonstrated in \figref{fig:bird}, AI models like the Segment Anything Model (SAM) can encode the original image {\rm (a)} into a compact sequence of tokens: 
{\tt [887, 3979, 349, 720, 2809, ..., 1326, 2598] with shape ([32] tokens)}.

When all tokens are accurately received, the generative AI model produces a very similar image shown in {\rm (b)} (MS-SSIM score: $0.598$.) 
However, if an error occurs in the least significant bit of the first token, while all other tokens remain correct, the resulting image is displayed in {\rm (c)} (MS-SSIM score: $0.541$,) revealing a noticeable alteration near the wing. 
In contrast, an error in the most significant bit of the first token produces the image shown in {\rm (d)}, significantly altering the depiction of the bird's head (MS-SSIM score: $0.539$.)
This example showcases the bit-level importance in token communications. 

In summary, the semantic importance of interest can be classified into the following categories:

\subsubsection{Segment-level  importance (SLI)} This measures how critical a semantic segment is to a specific task.

\subsubsection{Receiver-specific importance (RSI)} The importance of a segment varies depending on the needs of receivers.

\subsubsection{Token-level importance (TLI)} This measures the significance of individual tokens and categorizes them into three groups:
{\it a) highly critical tokens}, which are very sensitive to errors; {\it b) moderately robust tokens}, in which errors are detectable but not correctable; 
{\it c) highly robust tokens}, which can tolerate errors due to inherent redundancy or contextual dependencies, allowing for error correction without retransmission.

\subsubsection{Bit-level importance (BLI) in tokens} This measures the significance of bits within a token depends on their role in encoding information.

With this foundation, we will discuss importance-aware source-channel coding in Section \ref{sec3}.

\section{Importance-Aware Source-Channel Coding Strategies: A Discussion}\label{sec3}
Building on the classification of semantic importance, this section explores strategies for importance-aware source-channel coding. 
The goal is to align data compression, resource allocation, and error protection mechanisms with the varying levels of importance across semantic elements to ensure task-oriented efficiency and reliability in communications.

\subsection{SLI-Aware Coding}
Assume that the visual data object is partitioned into $L$ semantic segments, with each segment represented by a random variable $X_l,~_{l=1,...,L}$.
To support reconstructing the visual data object at the destination, a semantic map, denoted by the random variable $X_0$, is also transmitted.  
Mathematically, $X_l$ for $l=1,...,L$ are conditionally independent given $X_0$ \cite{bagheri2024c2}.
Each segment is then independently compressed according to its rate-distortion function, which is mathematically expressed as:
\begin{equation}\label{equ01}
R(D_l)=\min\{I(X_l; \widehat{X}_l)|\mathbb{E}[d(X_l, \widehat{X}_l)]\leq D_l\}
\end{equation} 
where $R(D_l)$ is the rate-distortion function, $I(X_l; \widehat{X}_l)$ is the mutual information between the original and decoded data ($\widehat{X}_l$), $d(X_l, \widehat{X}_l)$ is the distortion between $X_l$ and $\widehat{X}_l$, and $\mathbb{E}(\cdot)$ is the expectation.
For visual data (i.e., continuous-valued data), squared-error distortion (SED) is commonly employed as a measure \cite{Cover2006}:
\begin{equation}\label{equ02}
d(X_l, \widehat{X}_l)=\|\mathbf{x}_l-\widehat{\mathbf{x}}_l\|^2/N_l
\end{equation}
where $\mathbf{x}_l$ (or $\widehat{\mathbf{x}}_l$) represents a realization of the random variable $X_l$ (or $\widehat{X}_l$), with a vector length of $N_l$.
Consequently, the importance of $X_l$ is defined by its distortion tolerance $D_l$. 

When source coding is performed to maximize the allowable distortion for each semantic segment, the overall SED can be expressed as:
\begin{IEEEeqnarray}{ll}\label{rate-dist1}
\bar{d}(X, \widehat{X})&=\frac{\sum_{l=1}^Nd(X_l, \widehat{X}_l) N_l}{\sum_{l=1}^NN_l}\\
&\geq\frac{\sum_{l=1}^Nd_{\min}(X_l, \widehat{X}_l) N_l}{\sum_{l=1}^NN_l}=d_{\min}(X_l, \widehat{X}_l),
\end{IEEEeqnarray}
where 
\begin{equation}\label{d_min}
d_{\min}(X_l, \widehat{X}_l)\triangleq\min(d(X_l, \widehat{X}_l)),~ l=1,...,L.
\end{equation}
Physically, $d_{\min}(X_l, \widehat{X}_l)$ also represents the maximum SED allowable when source coding is applied to $X$ as a whole. 
This indicates that the SED is fundamentally constrained by the minimum distortion achievable for the individual semantic segments, $d(X_l, \widehat{X}_l)$.
Therefore, the difference between $\bar{d}(X, \widehat{X})$ and $d_{\min}(X_l, \widehat{X}_l)$, i.e.,
\begin{equation}\label{gain_1}
\Delta_l=\bar{d}(X, \widehat{X})-d_{\min}(X_l, \widehat{X}_l),
\end{equation}
indicates the gain of performing individual segment-level compression. 
On the other hand, the transmission of $X_0$ constitutes an additional overhead to the overall transmission rate.
Our experimental results show that $X_0$ can add around $10\%\sim30\%$ overhead onto the compressed data, depending on the complexity of the semantic map. 

\begin{figure*}[t!]
    \centering
    \includegraphics[width=0.7\textwidth]{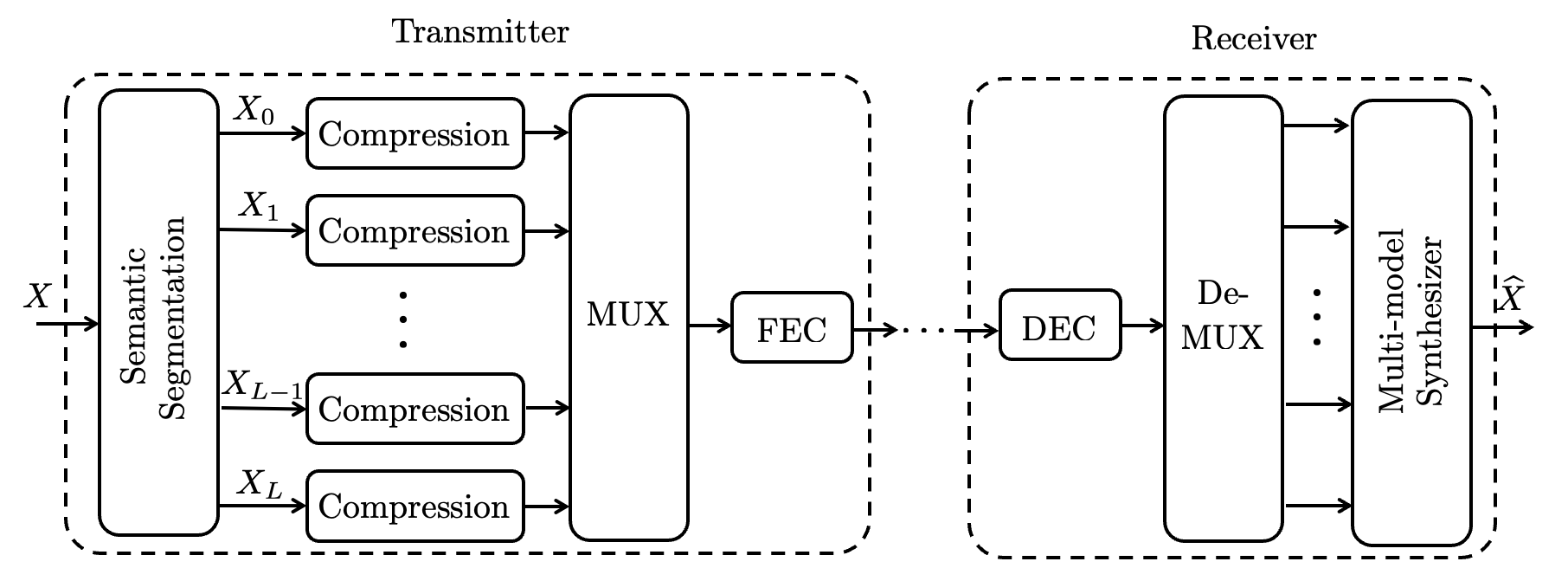} 
    \caption{Block diagram of the semantic segment-level importance-aware source-channel coding scheme. The compression block can incorporate AI-modal-based source encoding for efficient data representation.}
    \label{fig:SLI}
        \vspace{-10pt}
\end{figure*}

Notably, with this type of source coding strategy, all segments are highly compressed.
As a result, each segment is almost equally important to a task.
Channel coding and resource allocation designed for conventional sources remain optimal in this context.
Consequently, the SLI-aware source-channel coding approach reduces to SLI-aware source coding.
In conclusion, the SLI-aware source-channel coding scheme is illustrated in \figref{fig:SLI}.

\subsection{RSI-Aware Coding}
Consider a multiuser communication system involving a single transmitter and multiple receivers ($N$).
For the same semantic segment, receivers may have varying levels of interest based on their specific tasks.
In this scenario, the semantic importance of each segment should be indexed by the receiver, $n=1,...,N$.
Then, the SED function becomes:
\begin{equation}\label{equ08}
d_n(X_l, \widehat{X}_l)=\|\mathbf{x}_l-\widehat{\mathbf{x}}_{l,n}\|^2/N_{l}
\end{equation}
\begin{equation}\label{equ09}
\mathbb{E}[d_n(X_l, \widehat{X}_l)]\leq D_{l,n}.
\end{equation}
This presents a challenging task for source-channel coding and resource allocation, requiring adaptive strategies to address diverse receiver needs and optimize performance.

A potential solution can be developed based on the Slepian-Wolf principle of source coding, which addresses the efficient compression of correlated sources when partial information is available at the decoder \cite{Cover2006}. 
Specifically for the $l$-th segment, rate splitting is employed to divide the source $X_l$ into several parts ($K$): $X_{l,1}, X_{l,2}, ..., X_{l,K}$, 
where $X_{l,1}$ represents the part with the highest common interest among all receivers, 
and $X_{l,K}$ is the least significant part, tailored for receivers with minimal interest in certain details. 
Similar partitioning is also performed for other segments. 
Channel coding is then applied individually to each $X_{l,K}$, with optimally allocated resources corresponding to their importance. 
This method leverages the Slepian-Wolf coding framework, enabling receivers with distinct distortion requirements to receive only the necessary portions of the data, thereby optimizing both communication efficiency and resource allocation. By splitting the segment into multiple levels of importance, we ensure that each receiver receives only the data most relevant to their specific needs, enhancing performance in multiuser scenarios where semantic interests vary widely.

However, this approach introduces notable challenges in channel coding and resource allocation. 
Channel coding for the rate-splitting mechanism must ensure that the common and distinct portions of the data are recoverable under varying channel conditions for different receivers. 
For instance, the reliability of decoding $X_{l,1}$ (the part of highest common interest) must be guaranteed across all receivers, regardless of their individual channel conditions. 
This places stringent requirements on the design of codes that can operate effectively at multiple rates and levels of reliability simultaneously.
\begin{figure}[t!]
    \centering
    \includegraphics[width=0.43\textwidth]{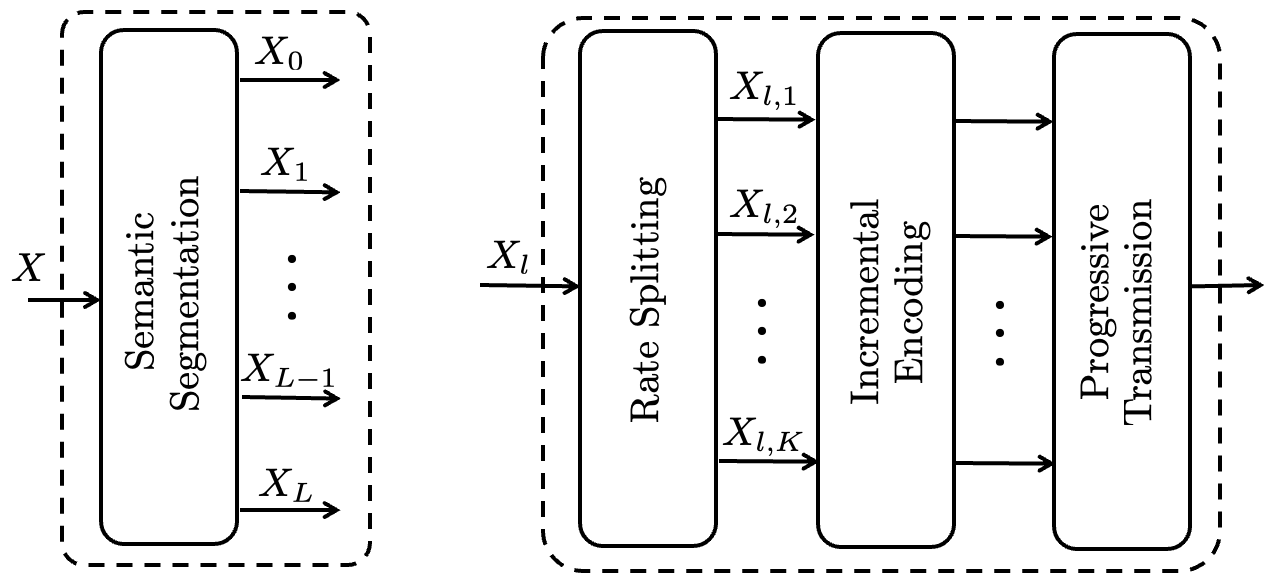} 
    \caption{Block diagram of the receiver specific importance-aware source-channel coding scheme (transmitter.)}
    \label{fig:RSI}
        \vspace{-10pt}
\end{figure}

Progressive transmission (e.g., in \cite{1096623}) offers a natural solution; see \figref{fig:RSI} for an illustrative overview.
Here, the segments are transmitted incrementally, beginning with the most important parts ($X_{l,1}$) and progressing to less significant parts ($X_{l,k}, k>1$), depending on the channel conditions and the interest levels of the receivers. 
This approach allows receivers with limited channel capacity or lower interest in details to stop decoding after receiving the essential parts, while receivers requiring higher fidelity or more specific information can continue to receive and decode additional layers of the transmission.

However, the combination of rate splitting and progressive transmission introduces challenges in channel coding and resource allocation. 
Channel coding must support incremental decoding, ensuring that each progressive layer of data is independently decodable without disrupting previously decoded layers. 
This demands sophisticated layered coding strategies that balance the reliability and decoding complexity across layers.
Resource allocation further requires adaptive optimization to prioritize the most critical layers while managing the limited resources available for subsequent layers. 
Power and bandwidth must be allocated dynamically, based on real-time feedback from receivers regarding channel conditions and task priorities. This ensures efficient use of resources while meeting the semantic distortion constraints for all receivers.

In summary, integrating rate splitting and progressive transmission can effectively address the diverse needs of multiuser systems, but it necessitates advanced coding schemes and real-time resource management to handle the complexity and variability of such scenarios.

\begin{figure}[t!]
    \centering
    \includegraphics[width=0.41\textwidth]{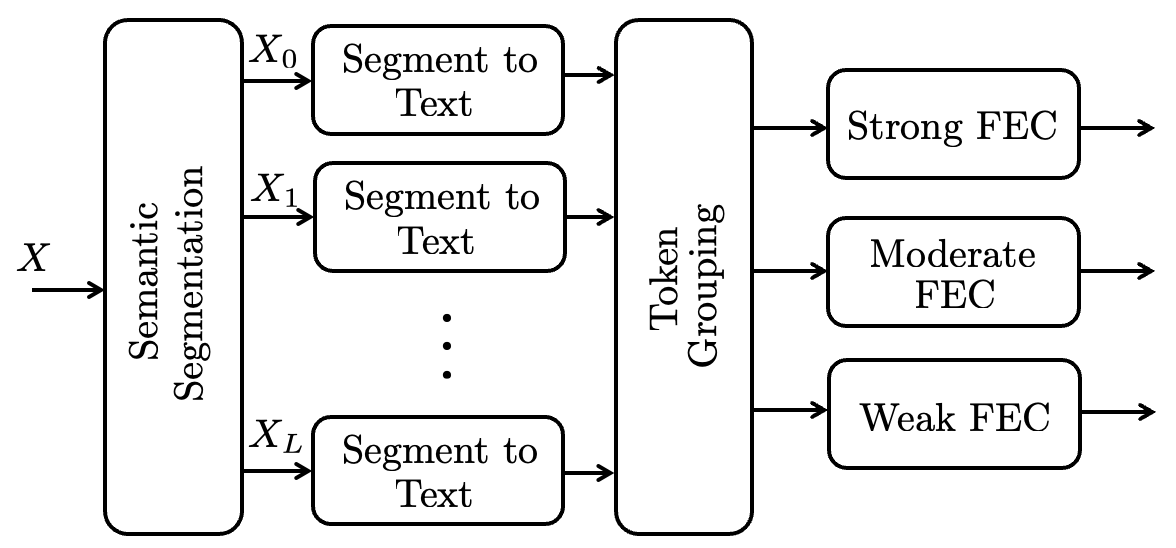} 
    \caption{Block diagram of the token-level importance-aware source-channel coding scheme (transmitter.)}
    \label{fig:TLI}
    \vspace{-15pt}
\end{figure}
\subsection{TLI-Aware Coding}
Designing a channel coding strategy for texture outputs with tokens classified by TLI requires prioritizing protection based on the sensitivity of token categories. 
The basic principle of TLI-aware coding is illustrated in \figref{fig:TLI}.
For highly critical tokens, which are highly sensitive to errors, robust error correction techniques like polar codes, low-density parity-check (LDPC) codes or turbo codes are employed. 
These codes provide strong forward error correction capabilities, ensuring that critical tokens are decoded with minimal error even under poor channel conditions. 
For moderately robust tokens, a less stringent error correction approach is sufficient since errors are detectable but not correctable. 
Hybrid automatic repeat request (HARQ) or adaptive error detection mechanisms can be applied here, providing an opportunity to retransmit only when errors exceed tolerable levels. 
Lastly, for highly robust tokens, minimal error correction is required due to inherent redundancy or contextual dependencies. 
Lightweight coding schemes like cyclic redundancy check (CRC) or systematic coding methods can be used, 
balancing computational complexity and resource allocation while tolerating minor errors without retransmission.

To address the challenge of short token lengths limiting the effectiveness of strong channel codes, grouping tokens of similar importance into larger blocks for joint channel coding offers a viable solution. 
By aggregating tokens within the same importance category (e.g., highly critical, moderately robust, or highly robust), we create longer message sequences that allow high-performance channel codes to approach their theoretical limits. 
For highly critical tokens, grouping ensures that LDPC or Polar codes achieve their full potential in error correction while maintaining strict reliability requirements. 
For moderately robust tokens, the aggregated block benefits from efficient HARQ strategies or lightweight error correction schemes optimized for mid-tier protection. 
Similarly, highly robust tokens grouped together can leverage simple parity checks or systematic encoding to maintain low overhead while tolerating minor errors without impacting the overall quality.

This approach not only enhances the coding performance but also simplifies resource allocation by treating groups of tokens as unified coding blocks. 
The grouping strategy can be dynamically adapted based on real-time system constraints, such as channel conditions or throughput requirements, ensuring a balance between computational efficiency and communication reliability. 
Such hierarchical grouping facilitates practical deployment in multiuser communication systems where semantic importance and channel conditions vary widely, aligning with the token-level importance framework.
\subsection{BLI-Aware Coding}
\begin{figure}[t!]
    \centering
    \includegraphics[width=0.41\textwidth]{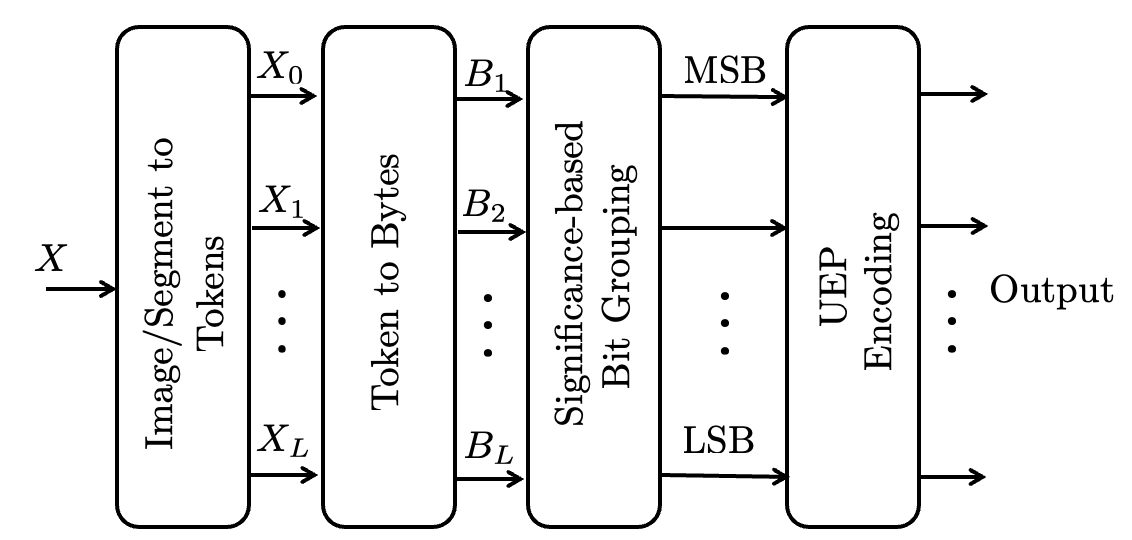} 
    \caption{Block diagram of the bit-level importance-aware source-channel coding scheme (transmitter.)}
    \label{fig:BLI}
        \vspace{-15pt}
\end{figure}
BLI-aware coding builds upon foundational concepts in unequal error protection (UEP) and adaptive resource allocation. 
Early research by Wolf and Cover \cite{Wolf1967,Cover1972} laid the groundwork for mechanisms that prioritize critical data by allocating resources based on its importance. 
These mechanisms dynamically adjust the level of error correction required for different segments of data, ensuring efficient and reliable communication.
\begin{figure}[t!]
    \centering
    \includegraphics[width=0.45\textwidth]{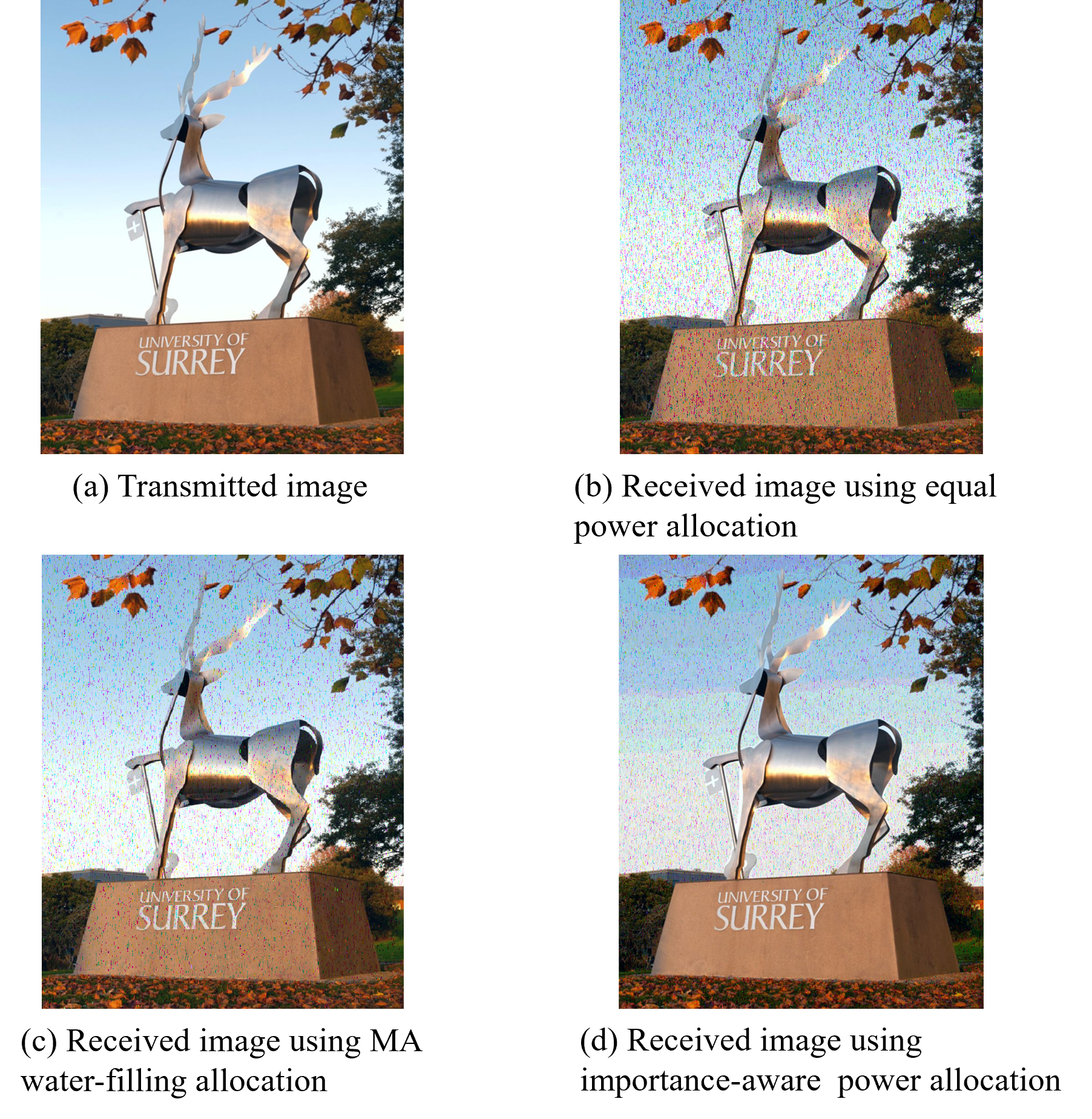} 
    \caption{Simulation results to demonstrate the effect of importance-aware optimum power allocation (Average SNR$=10$ dB).}
    \label{fig:IPA}
    \vspace{-10pt}
\end{figure}

A straightforward approach to BLI-aware coding involves grouping bits according to their importance; as depicted in \figref{fig:BLI}.
For instance, a stream can be formed by collecting all the most significant bits from tokens, while another stream consists of the least significant bits. Channel coding can then be applied separately to each stream, enabling tailored error correction strategies. 
This method aligns well with joint source-channel coding strategies, which optimize both the source encoding and channel transmission based on the importance of data segments.

Moreover, contemporary techniques such as adaptive modulation and coding enhance BLI-aware strategies by dynamically adjusting modulation schemes, transmission power, and error correction based on channel conditions and bit significance. 
For critical bits, strong error correction methods are employed, ensuring minimal error rates under adverse conditions. 
Conversely, for less critical bits, simpler coding schemes like CRC or systematic encoding are used to balance complexity and resource allocation. 
This tailored approach ensures that communication systems efficiently handle varying levels of bit importance, optimizing performance and reliability across diverse use cases in multimedia, IoT, and wireless communications.

To conclude our discussion, we present an experiment, with the results illustrated in \figref{fig:IPA}.
In this experiment, the image source is divided into three semantic segments: base, stag, and background. The semantic map is straightforward and assumed to be known during synthesis. The task-specific importance of these segments is: base ($49.75\%$), stag ($49.75\%$), and background ($0.5\%$), reflecting higher priority for base and stag. Each segment is transformed into a stream of tokens, with each token representing an $8$-bit byte. Bit-level importance is incorporated using the approach shown in \figref{fig:BLI}. As a result, the transmitter generates $24$ streams transmitted across $24$ orthogonal, independent sub-channels to the destination. For simplicity, each stream is encoded with a rate-$1/2$ convolutional code and modulated using $16$-QAM. Each stream is weighted based on both segment-level and bit-level importance during power allocation. Baselines for comparison include: (a) equal power allocation across all sub-channels and (b) conventional optimum power allocation based on SNRs at sub-channels. At an average-SNR of $10$ dB, our importance-aware approach clearly illustrates the base and stag segments more effectively than the baselines, with some trade-off in the less important background.

\section{Conclusion and Future Work}
In this paper, we have explored the integration of semantic importance into multi-modal task-oriented semantic communication systems. By utilizing GenAI to partition visual data into task-relevant semantic segments, we have developed a flexible framework that improves both accuracy and efficiency in data transmission. Our investigation of source and channel coding strategies highlights how semantic importance can be leveraged to dynamically allocate resources, ensuring high-fidelity transmission for critical information while enabling controlled degradation for less essential segments.

However, the source-channel coding strategies proposed in this work are still at an early stage of research. Rigorous mathematical analysis is needed to provide a comprehensive optimization of the proposed solutions. Additionally, extensive computer simulations are required to assess the effectiveness and performance benefits of semantic importance-aware approaches.

\section*{Acknowledgement}
This work was supported in part by the U.K. Department for Science, Innovation, and Technology under Project TUDOR (Towards Ubiquitous 3D Open Resilient Network).

\bibliographystyle{IEEEtran}
\bibliography{SemRef} 

\begin{thebibliography}{10}
\providecommand{\url}[1]{#1}
\csname url@samestyle\endcsname
\providecommand{\newblock}{\relax}
\providecommand{\bibinfo}[2]{#2}
\providecommand{\BIBentrySTDinterwordspacing}{\spaceskip=0pt\relax}
\providecommand{\BIBentryALTinterwordstretchfactor}{4}
\providecommand{\BIBentryALTinterwordspacing}{\spaceskip=\fontdimen2\font plus
\BIBentryALTinterwordstretchfactor\fontdimen3\font minus
  \fontdimen4\font\relax}
\providecommand{\BIBforeignlanguage}[2]{{%
\expandafter\ifx\csname l@#1\endcsname\relax
\typeout{** WARNING: IEEEtran.bst: No hyphenation pattern has been}%
\typeout{** loaded for the language `#1'. Using the pattern for}%
\typeout{** the default language instead.}%
\else
\language=\csname l@#1\endcsname
\fi
#2}}
\providecommand{\BIBdecl}{\relax}
\BIBdecl

\bibitem{JPEG2000}
C.~Christopoulos, A.~Skodras, and T.~Ebrahimi, ``The {JPEG2000} still image
  coding system: An overview,'' \emph{IEEE Trans. Consumer Electronics},
  vol.~46, no.~4, pp. 1103--1127, 2000.

\bibitem{kirillov2023segment}
A.~Kirillov \emph{et~al.}, ``Segment anything,'' in \emph{Proc. IEEE/CVF Int.
  Conf. Comput. Vis. (ICCV)}, Paris, France, Oct. 2023, pp. 4015--4026.

\bibitem{saharia2022}
C.~Saharia \emph{et~al.}, ``Photorealistic text-to-image diffusion models with
  deep language understanding,'' \emph{NeurIPS}, May 2022.

\bibitem{xu2024generative}
\BIBentryALTinterwordspacing
C.~Xu, M.~B. Mashhadi, Y.~Ma, R.~Tafazolli, and J.~Wang, ``Generative semantic
  communications with foundation models: Perception-error analysis and
  semantic-aware power allocation,'' 2024. [Online]. Available:
  \url{https://arxiv.org/abs/2411.04575}
\BIBentrySTDinterwordspacing

\bibitem{deletang2024language}
\BIBentryALTinterwordspacing
G.~Deletang \emph{et~al.}, ``Language modeling is compression,'' in \emph{The
  Twelfth International Conference on Learning Representations (ICLR)}, 2024.
  [Online]. Available: \url{https://openreview.net/forum?id=jznbgiynus}
\BIBentrySTDinterwordspacing

\bibitem{AK1963}
A.~Kolmogorov, ``On tables of random numbers,'' \emph{The Indian Journal of
  Statistics, Series A}, pp. 369--375, 1963.

\bibitem{AK1998}
------, ``On tables of random numbers,'' \emph{Theoretical Computer Science},
  pp. 387--395, 1998.

\bibitem{WenICCC2024}
W.~Tong, ``The {Kolmogorov} information theory based token communications:
  {From Weaver to AGI},'' Keynote presented at IEEE ICCC, Hangzhou, 2024.

\bibitem{liu2024semcom}
\BIBentryALTinterwordspacing
X.~Liu, M.~B. Mashhadi, L.~Qiao, Y.~Ma, R.~Tafazolli, and M.~Bennis,
  ``Diffusion-based generative multicasting with intent-aware semantic
  decomposition,'' 2024. [Online]. Available:
  \url{https://arxiv.org/abs/2411.02334}
\BIBentrySTDinterwordspacing

\bibitem{vaswani2017attention}
A.~Vaswani \emph{et~al.}, ``Attention is all you need,'' in \emph{Advances in
  Neural Information Processing Systems}, vol.~30.\hskip 1em plus 0.5em minus
  0.4em\relax Curran Associates, Inc., 2017, pp. 5998--6008.

\bibitem{bagheri2024c2}
\BIBentryALTinterwordspacing
A.~Bagheri, M.~Alinejad, K.~Bello, and A.~Akhondi-Asl, ``{C2P}: Featuring large
  language models with causal reasoning,'' 2024. [Online]. Available:
  \url{https://arxiv.org/abs/2407.18069}
\BIBentrySTDinterwordspacing

\bibitem{Cover2006}
T.~M. Cover and J.~A. Thomas, \emph{Elements of Information Theory}.\hskip 1em
  plus 0.5em minus 0.4em\relax USA: Wiley-Interscience, 2006.

\bibitem{1096623}
W.~Hofmann and D.~Troxel, ``Making progressive transmission adaptive,''
  \emph{IEEE Trans. Commun.}, vol.~34, no.~8, pp. 806--813, 1986.

\bibitem{Wolf1967}
B.~Masnick and J.~Wolf, ``On linear unequal error protection codes,''
  \emph{IEEE Trans. Inf. Theory}, vol.~13, no.~4, pp. 600--607, 1967.

\bibitem{Cover1972}
T.~Cover, ``Broadcast channels,'' \emph{IEEE Trans. Inf. Theory}, vol.~18,
  no.~1, pp. 2--14, 1972.

\end{thebibliography}
	
\end{document}